# Exceptional Superconducting Interfaces between Graphene and Atomic Chains in A15-MoRe

Hiroto Ogura[1, 2]*, Yuto Tsukidate[1, 2], Shun Kondo[2, 3], Zhuoqun Li[1, 2], Natsuki Sasaki[1, 2], Tsunenobu Sato[1, 2], Azusa Utsumi[2], Tomohiro Otsuka[1, 2, 4, 5, 6], Mitsuhiro Saito[3], Kazutoshi Inoue[2], Takashi Taniguchi[5], Kenji Watanabe[7], Yuichi Ikuhara[2, 3], and Toshiaki Kato[1, 2]*

[1]Graduate School of Engineering, Tohoku University, Sendai 980-8579, Japan
[2]Advanced Institute for Materials Research (AIMR), Tohoku University, Sendai 980-8577, Japan
[3]Institute of Engineering Innovation, School of Engineering, The University of Tokyo, Tokyo 113-8656, Japan
[4]Research Institute of Electrical Communication (RIEC), Tohoku University, Sendai 980-8577, Japan
[5]Research Center for Materials Nanoarchitectonics (MANA), National Institute for Materials Science (NIMS), Tsukuba 305-0047, Japan
[6]Center for Emergent Matter Science, RIKEN, Saitama 351-0198, Japan
[7]Research Center for Electronic and Optical Materials, National Institute for Materials Science (NIMS), Tsukuba 305-0044, Japan

*Email: hiroto.ogura.e8@tohoku.ac.jp; kato12@tohoku.ac.jp

**Realizing highly transparent and magnetic-field-resilient superconducting contacts is a critical bottleneck for the advancement of two-dimensional (2D) quantum devices. Molybdenum-rhenium (MoRe) alloys have been empirically favored as they uniquely yield ballistic, highly transparent contacts to graphene that survive in high-magnetic-field regimes; however, the microscopic origin of this anomalous transparency has remained a long-standing puzzle. Here, we resolve this paradox by revealing that pristine graphene/MoRe interfaces are governed by the classical A15 superconducting phase. Combining low-temperature transport, X-ray diffraction, and atomic-resolution electron microscopy, we demonstrate that as-deposited MoRe spontaneously forms the A15 phase in direct contact with graphene, whereas thermal annealing triggers a phase transition to a hexagonal close-packed (hcp) structure that severely degrades the Josephson transparency. Crucially, the one-dimensional (1D) transition-metal atomic chains unique to the A15 structure exhibit a near-perfect lattice match (mismatch < 1%) with graphene edges. First-principles calculations further prove that the *d*-orbitals of these historically renowned 1D chains possibly hybridize with graphene's Dirac cone near the K point. These findings not only uncover the structural and electronic origins of optimal MoRe contacts but also establish A15-phase superconductors as a geometrically ideal, universal platform for realizing robust quantum electronics based on 2D materials.**

Graphene-based Josephson junctions (GJJs) have emerged as a highly versatile platform for exploring exotic quantum phenomena and developing scalable quantum computing architectures[1-4]. By leveraging graphene's high mobility and unique band structure, these junctions have enabled the observation of a wide spectrum of quantum states, including Andreev bound states, chiral Andreev edge states, and quantum Hall–superconducting hybrid states[2,5-7]. Realizing these next-generation quantum technologies imposes remarkably stringent requirements on the superconducting electrodes: the contacts must be highly transparent to support ballistic transport, while simultaneously being robust against large magnetic fields. Elemental superconductors such as aluminum (Al) can form exceptionally clean interfaces with graphene, their vanishingly small critical magnetic fields fundamentally preclude their use in these next-generation quantum technologies. Consequently, discovering a superconducting contact that simultaneously guarantees ballistic transparency, high critical fields, and structural integrity remains a formidable challenge in 2D quantum electronics.

To circumvent these fundamental limitations, molybdenum-rhenium (MoRe) alloys have been empirically adopted as a pragmatic and highly successful electrode material in 2D device research. Figure 1a summarizes previous experimental studies by plotting the critical magnetic field of various superconducting electrodes against the critical current density of the corresponding GJJs[1-6,8-40]. As shown, MoRe exhibits a remarkably high critical magnetic field. More importantly, many MoRe-contacted GJJs consistently host purely ballistic Josephson transport, unequivocally proving the formation of highly transparent contacts[2,4-6,29,30]. To date, more than 30 experimental studies have reported the use of MoRe as superconducting electrodes, making it one of the most widely used superconducting electrode materials for GJJs. However, this empirical success presents a profound structural paradox: why does a sputtered binary alloy of MoRe, which is conventionally expected to be amorphous or structurally disordered at the interface, form a significantly more transparent contact to graphene than pure elemental metals? The microscopic origin of this anomalous interfacial transparency between MoRe and graphene has remained a long-standing mystery in the field.

In this work, we resolve this paradox by revealing that the pristine graphene/MoRe interface is not a random alloy, but rather an exquisite one-dimensional (1D) lattice-matched contact enabled by the classical A15 superconducting phase. Historically, A15-type intermetallic compounds (such as $Nb_3Sn$ and $V_3Si$) have dominated the field of high-critical temperature ($T_c$) superconductivity, yet their potential as contacts for low-dimensional materials has been entirely overlooked. By fabricating GJJs based on monolayer or bilayer graphene encapsulated in hexagonal boron nitride (hBN) with edge-contacted MoRe electrodes (Fig. 1b) and systematically tracking the structural evolution of the contacts, we demonstrate that the highest contact transparency is exclusively achieved in the as-deposited state. Utilizing atomic-resolution scanning transmission electron microscopy (STEM) and X-ray diffraction (XRD), we directly observe that the as-deposited MoRe film spontaneously forms the A15 phase in direct contact with the graphene edge, which subsequently transforms into a hexagonal close-packed (hcp) phase upon thermal annealing. Crucially, our lattice-matching analysis and first-principles calculations unveil that the unique 1D transition-metal atomic chains inherent to the A15 structure exhibit a nearly perfect lattice match (mismatch $< 1\%$) with graphene zigzag edges, providing compatibility for orbital hybridization. These findings not only clarify the microscopic origin of the highly transparent MoRe contacts, but also establish A15-phase superconductors as a universal and geometrically ideal platform for constructing robust quantum interfaces with 2D materials.

## Thermal degradation of ballistic contacts

To evaluate the quality of the superconducting contacts formed at the graphene/MoRe interface, we compared the low-temperature transport characteristics of GJJs annealed at different temperatures. Figure 2a shows a color map of the differential resistance ($dV/dI$) as a function of the current bias ($I$) and the gate-voltage offset from the charge-neutrality point ($\Delta V_g$) for the as-deposited device. In this device, a clear zero-resistance state is observed and the current range over which this state appears varies systematically with $\Delta V_g$. This behavior indicates that the critical current is controlled by the carrier density in graphene, consistent with the gate-tunable Josephson effect expected in GJJs[2,5]. The observation of a zero-resistance state without post-annealing suggests that transparent superconducting contacts can be formed at the graphene/MoRe interfaces without thermal treatment. Figures 2b–d show the corresponding differential resistance maps for devices annealed at 150, 500, and 900 °C, respectively. For the device annealed at 150 °C, both the zero-resistance state and the gate-tunable critical current are still observed, demonstrating that the superconducting proximity effect is preserved after moderate thermal treatment (Fig. 2b). In contrast, the devices annealed at 500 and 900 °C exhibit a pronounced degradation of the Josephson effect. In these devices, the differential resistance remains finite even in the low-bias-current regime (Fig. 2c,d). For the as-deposited and 150 °C annealed samples, the resistance drops to zero at low bias current, confirming the flow of a supercurrent (Fig. 2a,b,e-i,e-ii). The largest critical current is obtained in the as-deposited sample, whereas the 150 °C annealed sample shows a reduced critical current (Fig. 2e-i,e-ii). The samples annealed at 500 and 900 °C retain a finite differential resistance under the same gate condition and do not reach a zero-resistance state (Fig. 2e-iii,e-iv). These results demonstrate that MoRe electrodes form transparent superconducting contacts to graphene without annealing and the transparency of the graphene/MoRe interface is degraded by post-annealing.

## Spontaneous formation of the A15 phase

Figure 3a shows the XRD spectra of the as-deposited MoRe film and the films annealed at different temperatures. The calculated diffraction peak positions for the A15 and hcp phases of MoRe, as well as those for $MoO_2$, are also indicated (PDF#71-9799, PDF#71-9795, and PDF#78-1069). Here, the peaks observed at $2\theta \approx 35.6°$, 39.9°, 40.5°, and 42.8° are denoted as peaks A, B, C, and D, respectively. Comparison with the calculated peak positions indicates that peaks A and D nearly coincide with the diffraction peak positions associated with the A15 and hcp phases of MoRe, respectively. For the as-deposited MoRe film and the films annealed at 150–600 °C, peak A—originating from the A15 phase—was clearly observed. The intensity of Peak A decreased sharply after annealing at 700 °C and nearly disappeared at 800 and 900 °C. In contrast, peak D—originating from the hcp phase—appeared in the films annealed at 700–900 °C. These results indicate that high-temperature annealing induces a structural transformation of the MoRe film from the A15 phase to the hcp phase. For peaks B and C, a precise identification of their phase origins is challenging because both the A15 and hcp phases yield diffraction peaks at nearly identical $2\theta$ values. Nevertheless, the relative intensities of peaks B and C were found to change systematically with increasing annealing temperature. This variation is considered to reflect changes in the crystalline phases and/or crystallographic orientation of the MoRe film induced by annealing. Furthermore, as shown in Fig. 3a and Extended Data Fig. 1, several peaks associated with hcp-MoRe and $MoO_2$ were observed in the annealing-temperature range of 700–900 °C. The peaks at $2\theta \approx 37°$, 53°, and 60.4°, which are considered to originate from $MoO_2$, increased in intensity from 700 to 800 °C and subsequently decreased at 900 °C. In contrast, the peaks at $2\theta \approx$ 37.5°, 42.8°, and 56°, which are associated with the hcp phase, were strongly observed at 800 and 900 °C. These observations suggest that annealing at 700 and 800 °C promotes the formation of $MoO_2$, whereas the relative fraction of $MoO_2$ decreases at 900 °C, accompanied by an increase in the relative fraction of the hcp phase. To clarify the correlation between the structural changes of MoRe and the Josephson effect, the intensity ratios of Peaks A and D ($I_A/I_D$) and Peaks B and C ($I_B/I_C$) obtained from Fig. 3a, together with the minimum resistance of each device shown in Fig. 2, were plotted as functions of annealing temperature (Fig. 3b). The $I_A/I_D$ decreased sharply at temperatures of 700 °C and above, indicating a decrease in the A15 phase and the emergence of the hcp phase as the dominant phase (Fig. 3b-i). In addition, the $I_B/I_C$ decreased with increasing annealing temperature and dropped below 1 at 400 °C or higher (Fig. 3b-ii). The minimum resistance of the devices in Fig. 2 ceased to reach zero at approximately 500 °C, demonstrating that annealing degrades the Josephson effect (Fig. 3b-iii). This analysis suggests that the annealing-induced transformation of MoRe from the A15 phase to the hcp phase does not improve the Josephson effect but instead leads to its degradation. Moreover, the temperature at which the zero-resistance state disappears agrees well with the temperature at which the value of $I_B/I_C$ becomes below 1, suggesting that the structural changes in the MoRe film reflected in the relative intensities of peaks B and C are closely related to changes in the transparency of the graphene/MoRe interfaces. These XRD results demonstrate that the annealing-induced structural transformation of MoRe causes the degradation of the Josephson effect in GJJs. Consequently, using as-deposited MoRe electrodes predominantly composed of the A15 phase is essential for forming highly transparent graphene/MoRe contacts.

## Perfect 1D lattice matching at graphene edges

To further investigate the microstructural properties of the MoR and graphene/MoRe interdface, we performed cross-sectional STEM measurements. As a first step, we examined the cross-sectional structure of the as-deposited MoRe film. Figure 4a shows a high-angle annular dark field (HAADF)-STEM image and the corresponding energy-dispersive X-ray spectroscopy (EDS) maps of the as-deposited MoRe film on $SiO_2$. Mo and Re are distributed throughout the MoRe layer, whereas Si and O are localized in the $SiO_2$ substrate region. Figure 4b shows an annular bright-field (ABF) image of the same as-deposited MoRe film. The MoRe layer consists of an upper region composed of equiaxed grains and a lower region composed of columnar grains. Atomic-resolution HAADF-STEM observations of the lower columnar grain reveal a crystalline structure consistent with the A15 phase (Fig. 4c–e). In contrast, atomic-resolution imaging of the upper equiaxed grain shows an hcp phase (Extended Data Fig. 2). Similar cross-sectional STEM measurements were performed for MoRe films annealed at 500°C and 900°C (Extended Data Fig. 3 and Extended Data Fig. 4). For the MoRe film annealed at 500°C, equiaxed grains were observed not only in the upper region of the film but also near the $SiO_2$ surface, suggesting that annealing at 500°C promotes the formation of the hcp phase compared with the as-deposited film (Extended Data Fig. 3). After annealing at 900°C, the A15 phase was no longer observed, and an $MoO_x$ layer was formed between the upper and lower hcp regions (Extended Data Fig. 4). These results indicate that the A15 phase is characteristic of the as-deposited MoRe film and that high-temperature annealing substantially modifies the phase structure of the MoRe film.

While the aforementioned XRD and bulk STEM analyses establish the A15 phase as the dominant structure in the as-deposited MoRe films, they only provide macroscopic phase information. To unambiguously correlate this metastable phase with the highly transparent Josephson transport, it is imperative to obtain direct structural evidence at the exact junction interface. Therefore, we extracted the cross-sectional contact region directly from an actual functional device following its electrical characterization (Fig. 4f,g). The Josephson characteristics of this specific device used for the STEM observation are presented in Fig. 2a. Consistent with the as-deposited devices discussed earlier, it clearly exhibited a zero-resistance state and a gate-tunable critical current, confirming the presence of a robust, high-quality supercurrent prior to the structural analysis.

Figure 4f shows the low-angle annular dark-field (LAADF)-STEM image of this precise as-deposited graphene/MoRe interface. Strikingly, the MoRe electrode maintains the identical columnar grain structure characteristic of the A15 phase (as observed in Fig. 4b–e) perfectly down to the junction boundary. In addition, the LAADF-STEM image shows a thin interfacial contrast between the top and bottom hBN crystals, which is attributed to the graphene layer (Fig. 4g). This direct atomic-scale visualization definitively proves that the metastable A15 phase is not just present in the bulk film, but is in direct physical contact with the graphene edge, thereby acting as the true origin of the exceptional superconducting contact.

Our experimental observations have established that the metastable A15 phase in as-deposited MoRe is fundamentally responsible for the highly transparent superconducting transport into graphene. To uncover the microscopic mechanism underlying why this specific A15 structure serves as such an exceptional contact, we first focused on the lattice matching in real space. Notably, we discovered that the interatomic spacing of the one-dimensional (1D) atomic chains inherent to the A15 structure exhibits a nearly perfect match—with less than 1% mismatch—to the lattice constant of the graphene zigzag edges. Figures 5a–d schematically illustrate the characteristic interatomic spacings in the A15 and hcp phases of MoRe. Among the possible crystallographic directions, we selected atomic rows that form 1D chains with well-defined long-range periodicity. Figure 5e summarizes the calculated minimum lattice mismatch between each MoRe atomic row and the graphene edge. The minimum lattice mismatch $\delta_{min}$ is defined as follows[41]:

$$\delta_{min} = \min_{n,m} \left| \frac{n d_{MoRe} - m d_{Gr}}{m d_{Gr}} \right| \ (n, m = 1, 2)$$

where $d_{MoRe}$ and $d_{Gr}$ denote the atomic distances along the MoRe atomic chains (Fig. 5a–d) and the graphene edges (inset of Fig. 5e), respectively. The lattice parameters used in the calculations are a = b = c = 4.95 Å for the A15 phase[42], and a = b = 2.771 Å, c = 4.486 Å for the hcp phase[43]. The graphene lattice constant is taken to be 2.46 Å.

As plotted in Fig. 5e, the lattice mismatch between the 1D chains in the A15 (100) plane and the graphene zigzag edge is remarkably low at 0.6%. Furthermore, for the armchair edge, the diagonal atomic arrangement in the A15 (110) plane also demonstrates excellent registry. These representative lattice-matching configurations are schematically illustrated in Fig. 5f. In stark contrast, none of the possible atomic configurations in the hcp phase achieve a mismatch below 1% for any crystal plane. This remarkable real-space structural commensurability constitutes one of the key factors of the highly transparent superconducting contacts formed at the A15-MoRe/graphene interface.

## Orbital hybridization in reciprocal space

While the exceptional real-space structural commensurability provides the geometric foundation for the highly transparent interface, a comprehensive understanding of the superconducting contact requires assessing energy and momentum compatibility in reciprocal space. Carrier transport in graphene is primarily governed by the electronic states near the Dirac point (K point). Given that these K-point states are intimately associated with zigzag edges, which perfectly match the 1D chains in A15-MoRe, it is crucial to examine whether the electronic states of the A15-MoRe 1D chains can hybridize with those of graphene near the Dirac point.

To address this, we performed DFT calculations using a structural model of A15-MoRe. It is difficult to precisely determine the composition ratio of Mo and Re in the MoRe electrode in contact with graphene. Therefore, to simplify the calculations, we assumed that each one-dimensional (1D) chain identical Mo or Re atoms and carried out calculations for two distinct compositions, $Mo_3Re$ and $MoRe_3$ (Fig. 6a).

The calculated band structures of $Mo_3Re$ and $MoRe_3$ exhibit pronounced differences (Fig. 6b). We focus on the electronic states near the K point, within a wave-vector radius of $0.03\ \text{Å}^{-1}$, corresponding to the Fermi surface radius at $\Delta V_g = 40\ \text{V}$ in the devices shown in Figs. 2a, 4f, and 4g. In $Mo_3Re$, electronic states with a large $d$-orbital weight of the 1D chain (6c site) are located at $E - E_F \approx 650\ \text{meV}$. In contrast, in $MoRe_3$, the states with a large $d$-orbital weight are located at $E - E_F \approx 120\ \text{meV}$, which is considerably closer to the Fermi level than those in $Mo_3Re$. This result indicates that the $d$-orbital states in $MoRe_3$ are more favorably aligned in energy for coupling with the electronic states of graphene near the Fermi level. Furthermore, upon formation of the graphene/MoRe interface, the electronic states are modified by charge transfer and hybridization with the graphene $\pi$-orbitals, which may further enhance the $d$-orbital contribution to the states near the Fermi level. Therefore, particularly in $MoRe_3$, strong electronic coupling between the graphene Dirac states and the MoRe $d$-orbital states may develop. Such orbital coupling could serve as a key mechanism for enhancing the interfacial transparency between graphene and MoRe, thereby promoting the superconducting proximity effect.

To further examine the momentum dependence of these electronic states in the $MoRe_3$, we calculated two-dimensional maps of the $d$-orbital weight (Fig. 6c) and $E - E_F$ (Fig. 6d), respectively. In the vicinity of the folded graphene K point, the electronic states exhibit a large $d$-orbital weight and a relatively small $E - E_F$. This substantial $d$-orbital contribution near the Fermi level at wave vectors relevant to the graphene Dirac states further supports the possibility of strong electronic coupling between graphene and MoRe.

Historically, the *d*-electron states associated with these orthogonal 1D chains in A15 compounds have been considered to contribute to their high superconducting transition temperatures[44]. Our calculations suggest that *d*-orbital states from these 1D chains can couple to the Dirac states of graphene, providing a plausible microscopic framework for the formation of highly transparent superconducting contacts and thereby bridging classical high-field superconductivity with modern 2D quantum devices.

## Conclusions

In conclusion, we have elucidated the microscopic origin of the highly transparent superconducting contacts in graphene/MoRe Josephson junctions. Our results demonstrate that the exceptional ballistic transport empirically observed in these devices does not originate from a disordered alloy interface, but rather from the spontaneous formation of the classical A15 superconducting phase at the as-deposited contact. Combined structural and electrical characterizations clearly establish that thermal annealing destroys this advantageous A15 phase in favor of an hcp phase, directly causing the degradation of the Josephson effect.

The superiority of the A15 phase as a contact material for 2D systems is rooted in its unprecedented compatibility with graphene in both real and reciprocal spaces. Geometrically, the 1D atomic chains intrinsic to the A15 structure provide an exceptional lattice match (less than 1% mismatch) to the 1D graphene edges. Electronically, our DFT calculations reveal that the *d*-orbitals originating from these 1D chains—historically recognized as the primary drivers of robust superconductivity in A15 compounds—possibly hybridize with the graphene Dirac states.

By unraveling this long-standing empirical mystery, our work bridges classical high-magnetic-field superconductivity with modern low-dimensional physics. Beyond MoRe, these fundamental insights establish the broader family of A15-phase superconductors as a universal, geometrically ideal platform for engineering highly transparent and robust contacts. This design principle overcomes the inherent limitations of conventional elemental superconductors, paving the way for the realization of next-generation topological and quantum computing architectures.

## Methods

### Preparation of the hBN/graphene/hBN heterostructures

Monolayer or bilayer graphene and hBN flakes were prepared by mechanical exfoliation using Scotch tape. The layer number of graphene was determined by Raman spectroscopy using a 532 nm excitation laser (LabRAM HR-800, HORIBA). Monolayer and bilayer graphene were identified based on the intensity ratio of the 2D and G peaks and the line shape of the 2D peak. Representative Raman spectra of monolayer and bilayer graphene are shown in Extended Data Fig. 5. The hBN/graphene/hBN heterostructures were fabricated on $SiO_2$ (280 nm)/Si substrates by a dry-transfer method[45]. The thermoplastic resin used for the dry-transfer was prepared by mixing acrylic resin powder (BR-60, Mitsubishi Chemical) and anisole (Sigma-Aldrich) at a weight ratio of 6:4. The mixture was left at room temperature for 1 day to completely dissolve the powder to anisole. A droplet of the resin solution was placed on a glass slide using a toothpick and baked on a hot plate at 165 °C for 5 min to evaporate the anisole. For the pickup process, the resin droplet on the glass slide was brought into contact with an exfoliated flake at room temperature. The stage was then heated to 110 °C and subsequently cooled to below 50 °C before the flake was picked up. This pickup sequence was repeated three times to assemble the hBN/graphene/hBN stack. The resulting hBN/graphene/hBN stack was then brought into contact with the target substrate at room temperature and heated to 180 °C. While the temperature was maintained at 180 °C, the glass slide was lifted, leaving both the hBN/graphene/hBN stack and the resin on the substrate. Finally, the substrate was immersed in chloroform for 1 min to remove the resin, leaving the hBN/graphene/hBN heterostructure on the substrate.

**Device fabrication and electrical measurements**

After preparation of the hBN/graphene/hBN heterostructures, device patterns with a channel length of 400 nm were defined by electron-beam lithography. A bilayer resist structure was prepared by first spin-coating of PMMA 495 A4 at 2000 rpm for 1 min and baking on a hot plate at 180 °C for 30 s, followed by spin-coating of PMMA 950 A4 at 2000 rpm for 1 min and baking at 180 °C for 60 s. The device patterns were written using an electron-beam lithography system (ELS-7500X, ELIONIX) at an exposure dose of 500 μC/cm$^2$. Following the exposure, the resist was developed in MIBK/IPA (1:3) for 1 min and rinsed in IPA for 30 s. To expose the graphene layer prior to contact deposition, the top hBN layer was etched using $CF_4$ plasma. The samples were loaded into an etching system (SEDE-PHL, Meiwafosis Co., Ltd.) and evacuated to a pressure below 2 Pa. $CF_4$ gas was then introduced to maintain a working pressure of 20 Pa, and the $CF_4$ plasma irradiation was performed for 70 s at a discharge current of ~20 mA. Immediately after the etching process, a 100 nm-thick MoRe film was deposited by dc magnetron sputtering (QAM-4-S, ULVAC). The samples were placed in the load-lock chamber, which was evacuated to below $5 \times 10^{-4}$ Pa before transferring them to the process chamber. The base pressure of the process chamber was maintained below $2 \times 10^{-5}$ Pa. Ar gas was then introduced into the process chamber to reach a working pressure of 0.1 Pa. The MoRe layer was deposited at a plasma power of 20 W with a deposition rate of 2 nm/min for 50 min. The lift-off process was carried out by immersing the sample in hot acetone at 60 °C on a hot plate. To prepare the annealed devices, thermal annealing was performed following the lift-off process. The annealing chamber was evacuated to a base pressure of ~0.2 Pa, and the samples were annealed at temperatures ranging from 150 °C to 900 °C for 30 min. Electrical transport measurements were performed at 4.2 K. The devices were mounted on a dipstick and cooled down by immersion in liquid helium. The electrical resistance was measured using a lock-in technique with a lock-in module (LI5502, NF) at an excitation frequency of 170 Hz and an excitation current of 10 nA. To detect small signals, the measured voltage was amplified by a factor of 100 using a low-noise differential FET amplifier (SA-440F5, NF). The gate voltage was applied using a modular DC source/monitor (HP 4142B, Keysight).

## Structural characterizations

XRD measurements of the MoRe films were performed using Cu Kα1 radiation with a wavelength of 0.15406 nm (SmartLab 9KAB, Rigaku). For STEM observations, samples were prepared by focused ion beam (FIB) (Helios 5, FEI). The STEM characterizations were conducted using

ARM-200F (JEOL Ltd.) operated at 200 kV, equipped with an aberration corrector and double SDD-EDS detectors. The convergence semi-angle was 22 mrad, and the collection semi-angle was 90–175 mrad for HAADF, 40–160 mrad for LAADF, and 12–23 mrad for ABF imaging.

## First-principles calculations

First-principles calculations were performed using the projector augmented-wave (PAW) method within the framework of density functional theory (DFT), as implemented in the Vienna *ab initio* Simulation Package (VASP)[46-48]. The PAW formalism was used to describe the electron–ion interaction[49,50]. The exchange–correlation interaction was treated within the generalized-gradient approximation using the Perdew–Burke–Ernzerhof functional[51]. The plane-wave cutoff energy was set to 520 eV, and the electronic self-consistency criterion was $10^{-7}$ eV. The real-space projection approximation was disabled, and non-spherical contributions inside the PAW spheres were included. Metallic occupations were treated using the Methfessel–Paxton smearing scheme[52] with a small smearing width. The A15-type MoRe structure was modeled using an eight-atom primitive cubic cell containing six Mo atoms and two Re atoms. A15 compounds are characterized by three mutually orthogonal transition-metal chains, which have long been discussed in relation to their electronic instabilities and superconducting properties[44,53]. The lattice constant used in the present electronic-structure analysis was $a = 4.95$ Å. The corresponding reciprocal-lattice magnitude is $2\pi/a \simeq 12.69\ \mathrm{nm}^{-1}$. By using this unit cell, k-point convergence is systematically examined and the $\Gamma$-centered $36 \times 36 \times 36$ $k$-point mesh is applied for the self-consistent and non-self-consistent calculations. To assess the in-plane momentum compatibility between graphene and A15-MoRe, the graphene K point was folded into the first Brillouin zone of the A15 reciprocal lattice.

## Data availability

All data supporting the results of this study are available within the article. Additional data are available from the corresponding author upon request.

## Acknowledgements

This work was supported by JSPS (KAKENHI JP22H05441, JP23H00097, JP23K17756, JP23K19193, JP24H01165, JP24K17584, and JP26K22456), JST-CREST (JPMJCR23A2 and JPMJCR24A5), JST-SPRING (JPMJSP2114) the Murata Science Foundation, KONDO-ZAIDAN, MATSUDA-ZAIDAN, World Premier International Research Center Initiative (WPI), MEXT, Japan, the Cooperative Research Project Program of the Research Institute of Electrical Communication (RIEC), Tohoku University, research equipment sharing system, Tohoku University (1005 Laser Raman Microscope, 1007 Sputtering System, 529 Multipurpose X-ray Diffraction Integrated System, 471 ULVAC Multipurpose RIE System, and 735 Elionix 50 kV EB Lithography System), Tohoku University Microfabrication Platform in MEXT Advanced Research Infrastructure for Materials and Nanotechnology in Japan (JPMXP1223TU0100, JPMXP1224TU0013, JPMXP1225TU0022, and JPMXP1226TU0008) and support system for young researchers to use research equipment, instruments and devices in Tohoku University. We thank as well RIEC Fundamental Technology Center and the Laboratory for Nanoelectronics and Spintronics for technical support.

## Author contributions

H.O and T.K. developed the concept and supervised the project. H.O., Y.T., Z.L., N.T., T.S., A.U, and T.O. performed the device fabrication and electrical measurements of GJJs. Y.T., S.K., K.I., and Y.I. performed XRD measurements. S.K., M.S., K.I., and Y.I. conducted the STEM observations. H.O. and T.K. analyzed the lattice matching between graphene and MoRe. S.K., K.I., and Y.I. carried out the DFT calculations. K.W. and T.T. grew the bulk hBN crystals. H.O. and T.K. prepared the figures and wrote the manuscript. All authors discussed the results and commented on the paper.

## Competing interests

The authors declare no competing interests.

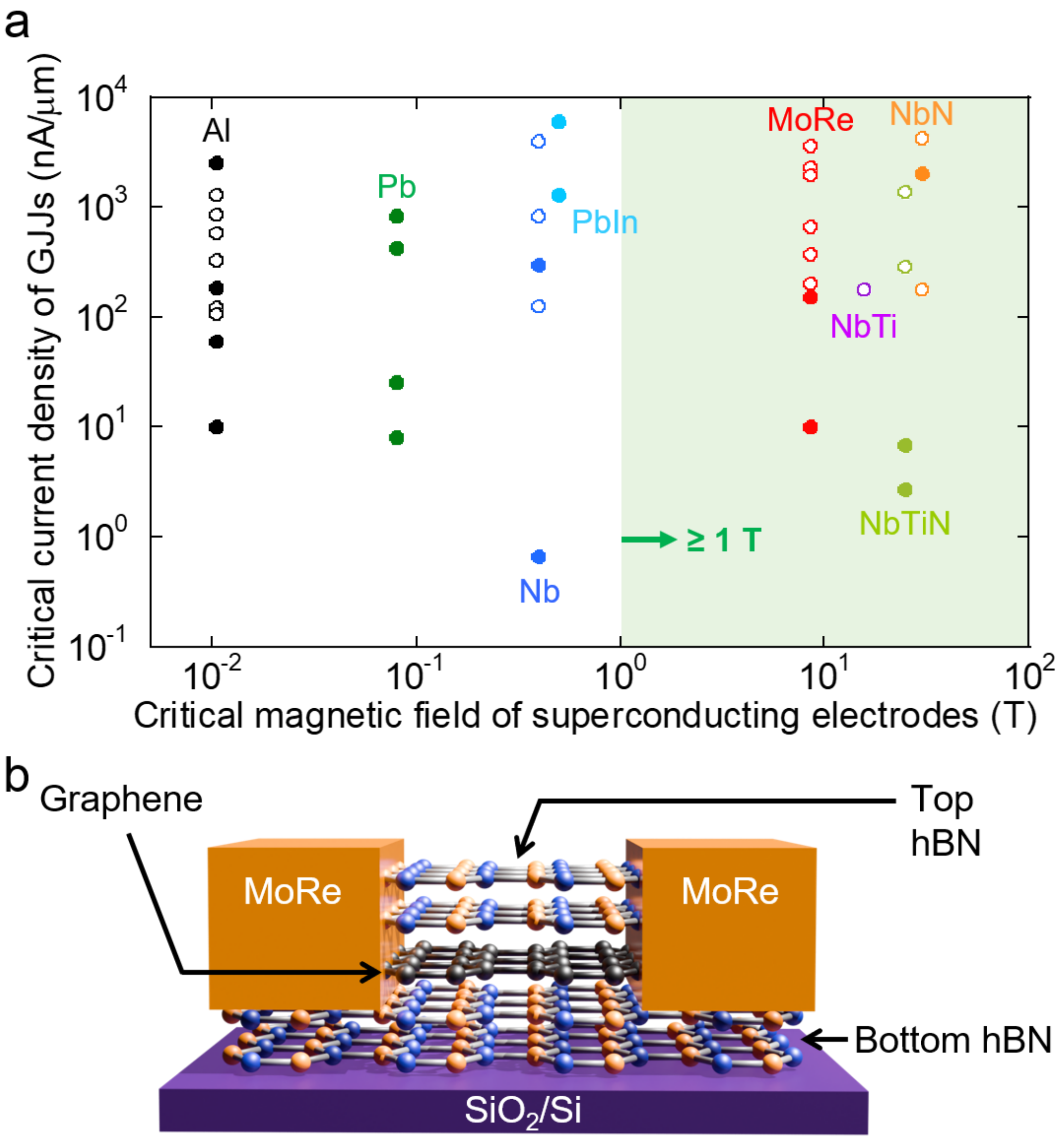

**Fig. 1 | Overview of the performance landscape and device structure of GJJs. a** The critical magnetic field of superconducting electrodes and critical current density of the corresponding GJJs obtained from recent reports[1-6,8-40]. White-filled circles indicate GJJs for which ballistic transport has been reported. The region of 1 T and above is highlighted by the muted yellow-green shading. **b** Schemataic of typical GJJ based on hBN/graphene/hBN heterostructure and MoRe electrodes.

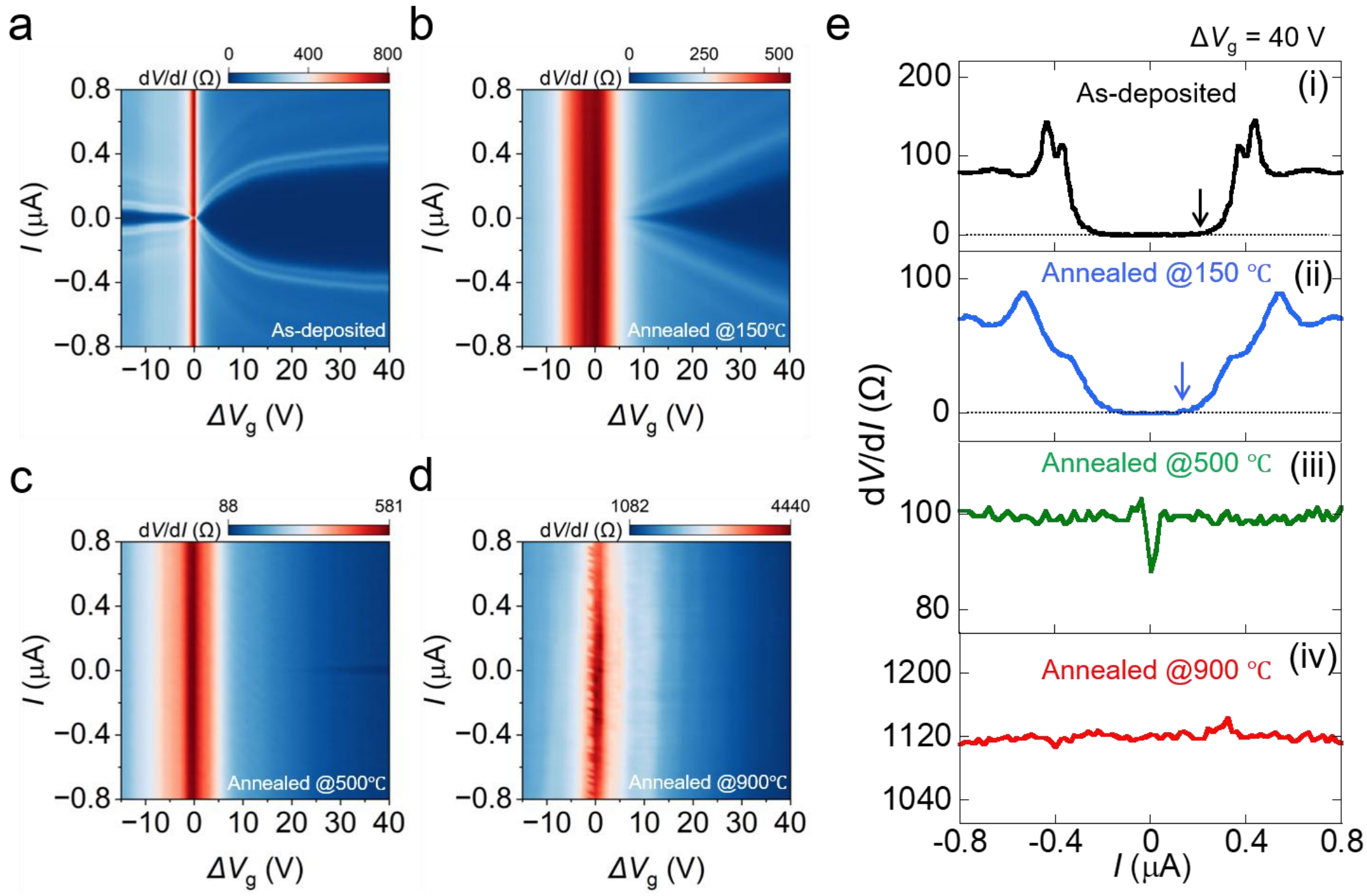

**Fig. 2 | Thermal annealing of GJJs. a–d** Differential resistance d$V$/d$I$ as a function of $\Delta V_g$ and $I$ for the as-deposited device (**a**) and devices annealed at 150 (**b**), 500 (**c**), and 900 °C (**d**). **e** Linecuts of the resistance maps (**a–d**) at $\Delta V_g$ = 40 V. Panels (i)–(iv) correspond to **a–d**, respectively, with arrows in (i) and (ii) indicating the critical current.

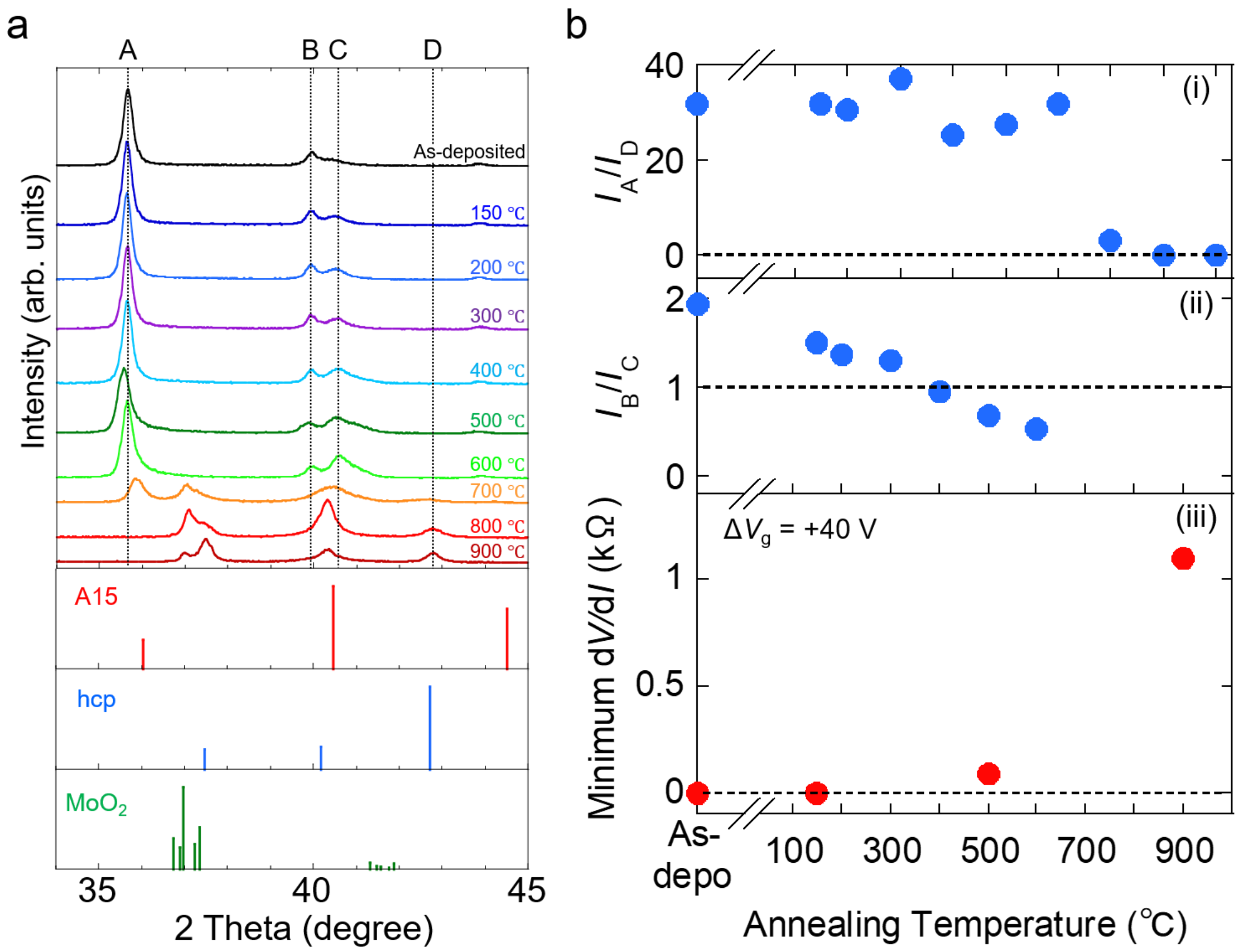

**Fig. 3 | Crystal phases of MoRe. a** XRD spectra of as-deposited and annealed MoRe films with calculated peak positions of A15-MoRe (PDF#71-9799), hcp-MoRe (PDF#71-9795), and $MoO_2$ (PDF#78-1069). **b** Annealing temperature dependence of the (i) $I_A/I_D$ and (ii) $I_B/I_C$ peak intensity ratios obtained from **a,** and (iii) the minimum d$V$/d$I$ at $\Delta V_g$ = 40 V from **2b**.

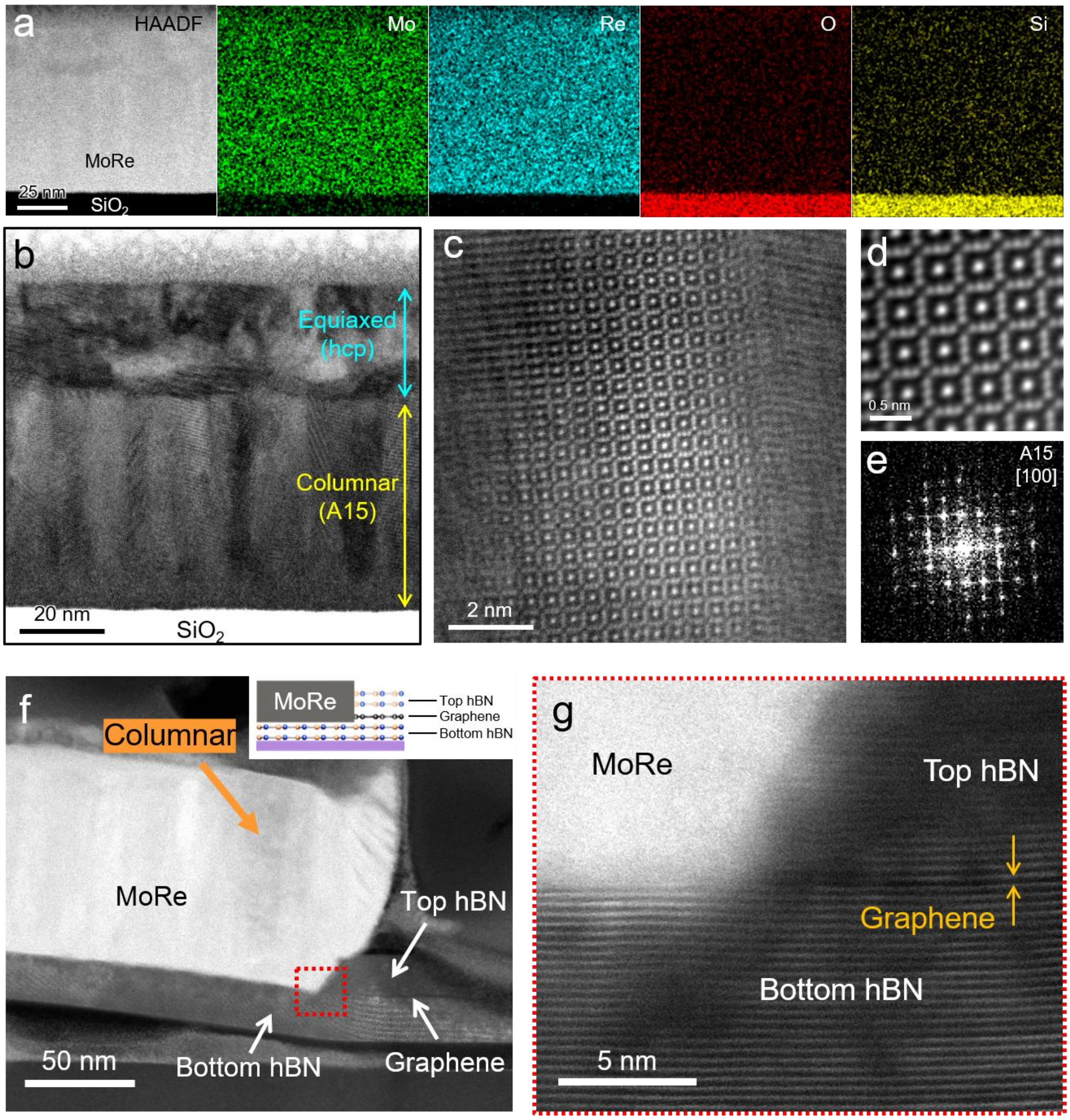

**Fig. 4 | STEM observations of the as-deposited MoRe and graphene/MoRe interface.** Panels **a–e** correspond to the as-deposited MoRe film, whereas **f** and **g** show cross-sections prepared from the channel region of the device in **2a**. **a** Cross-sectional HAADF-STEM image of the as-deposited MoRe on $SiO_2$ and EDS mapping images for Mo, Re, O, and Si. **b** ABF image of the as-deposited MoRe. **c** Atomic-resolution HAADF-STEM image of the lower columnar grain in **b**. **d** Position-averaged HAADF-STEM image obtained from **c**. The images of **c** and **d** were processed with the average background subtraction filter (ABSF) for the visibility of atomic columns. **e** Fast Fourier transform (FFT) pattern of the lower columnar grain in **b**. **f** ABF image of the as-deposited graphene/MoRe interface. The inset shows the schematic image of the interface. **g** High-resolution LAADF-STEM image of the as-deposited graphene/MoRe interface indicated by the dotted red square in **f**.

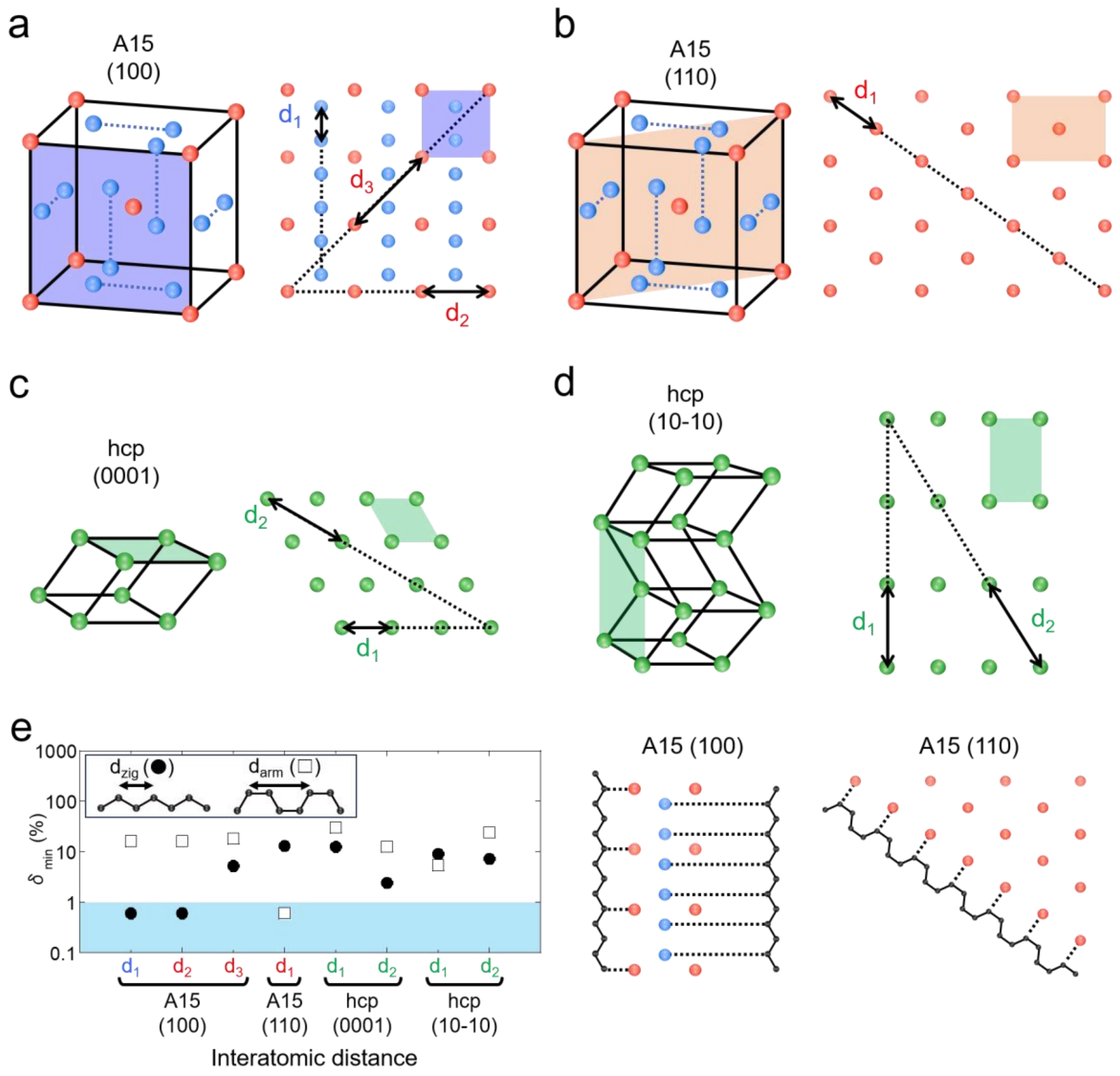

**Fig. 5 | Lattice matching between MoRe and graphene edges. a–d** Schematic illustrations of the crystal structures and characteristic interatomic spacings of the A15 (100) (**a**), A15 (110) (**b**), hcp (0001) (**c**), and hcp (10-10) (**d**) planes of MoRe. **e** Summary of the minimum lattice mismatches between the MoRe planes and graphene edges. Inset shows the atomic distances along graphene edges. **f** Schematic illustrations of the combinations in **e** with lattice mismatches of less than 1%.

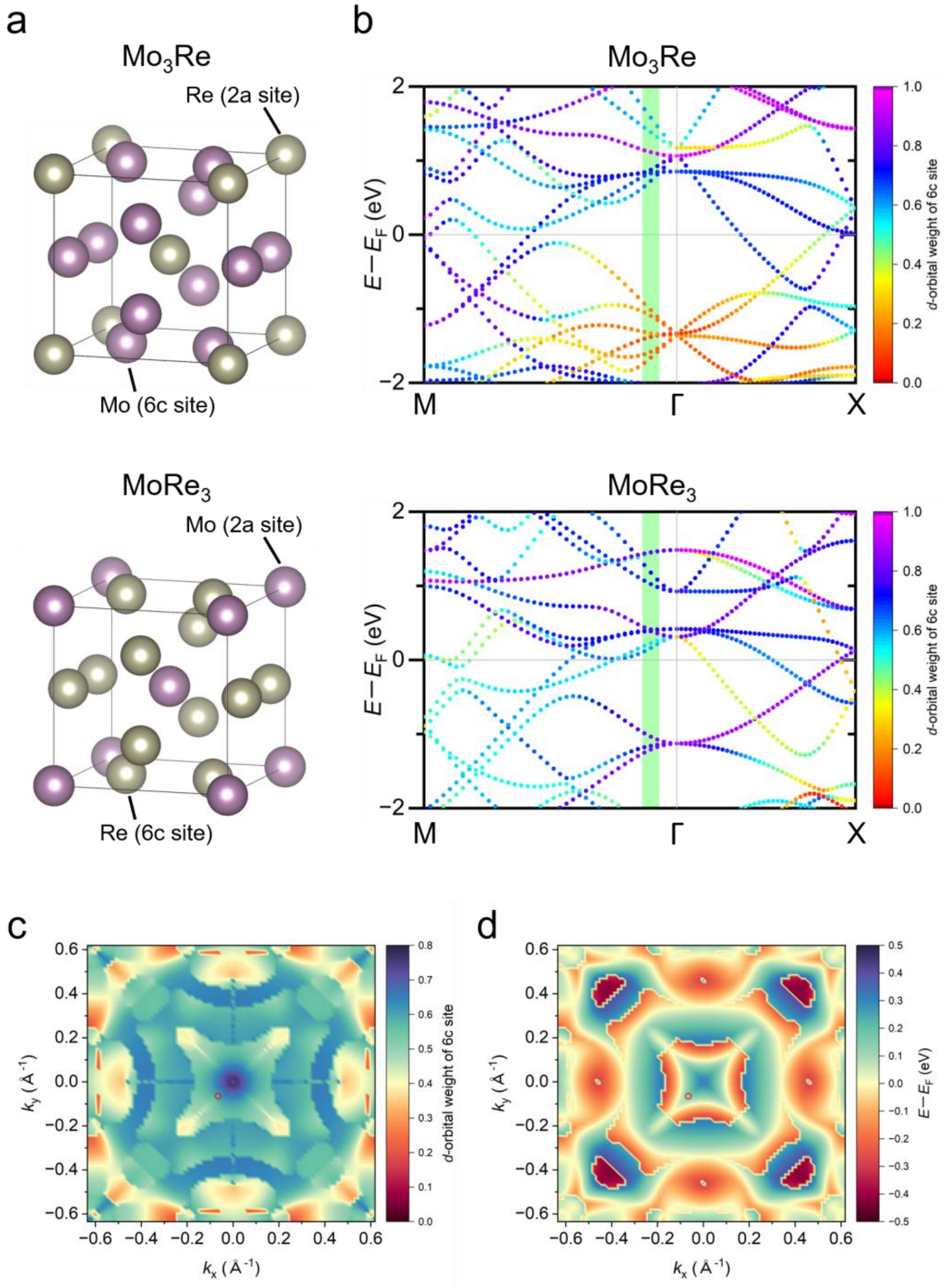

**Fig. 6 | Electronic-state compatibility between graphene and A15-phase MoRe. a** Unit cells of the A15-phase $Mo_3Re$ and $MoRe_3$ used in the present calculations. **b** Electronic band structures of the A15-phase $Mo_3Re$ and $MoRe_3$ along the M–Γ–X direction in reciprocal-space. The light green shaded area represents the range of $\pm 0.03$ Å$^{-1}$ from the folded graphene K point, which corresponds to the Fermi surface radius within the Dirac cone at $\Delta V_g = 40$ V for the device shown in **2a**, **4f**, and **4g**. **c,d** Momentum-resolved map of $d$-orbital weight from 6c-site (c) and $E - E_F$ (d) for the A15-phase $MoRe_3$. The red open circles in **c,d** indicate the position of the folded graphene K point.

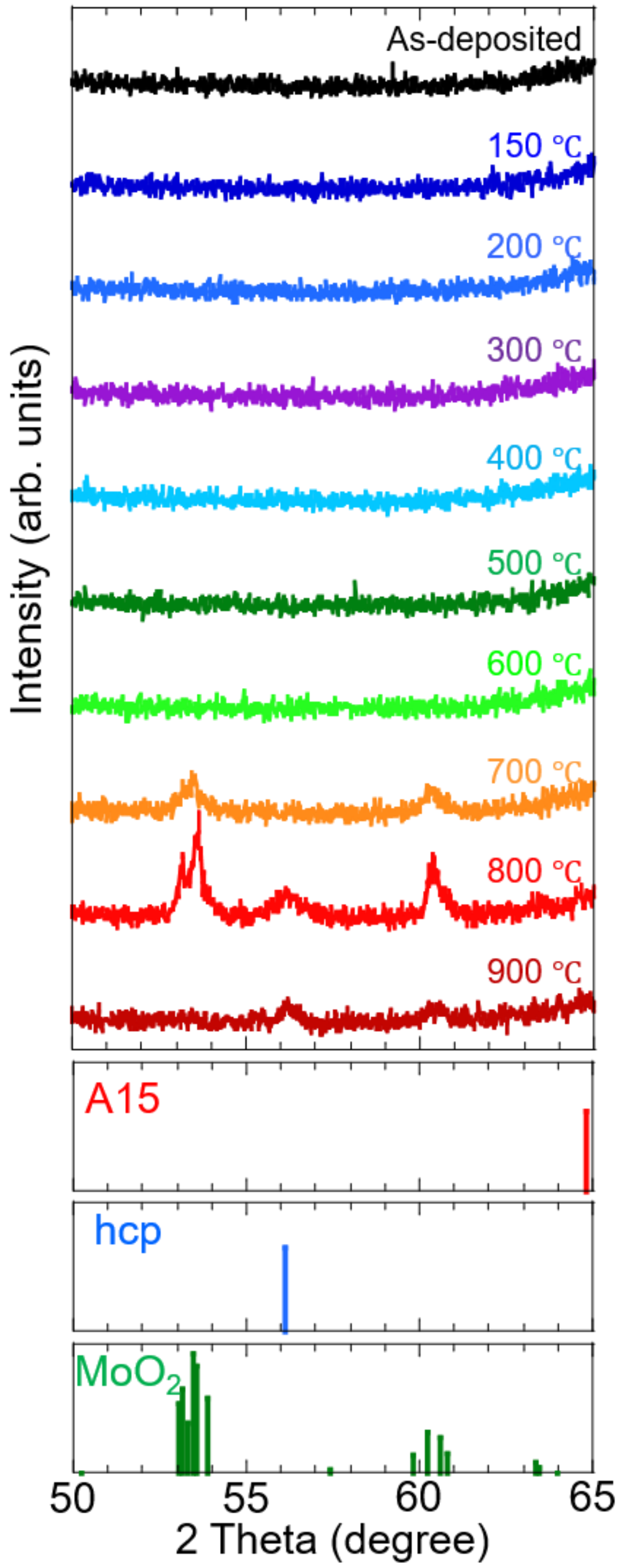

**Extended Data Fig. 1 | XRD spectra of as-deposited and annealed MoRe films at high diffraction angles with calculated peak positions of A15-MoRe (PDF#71-9799), hcp-MoRe (PDF#71-9795) and $MoO_2$ (PDF#78-1069).**

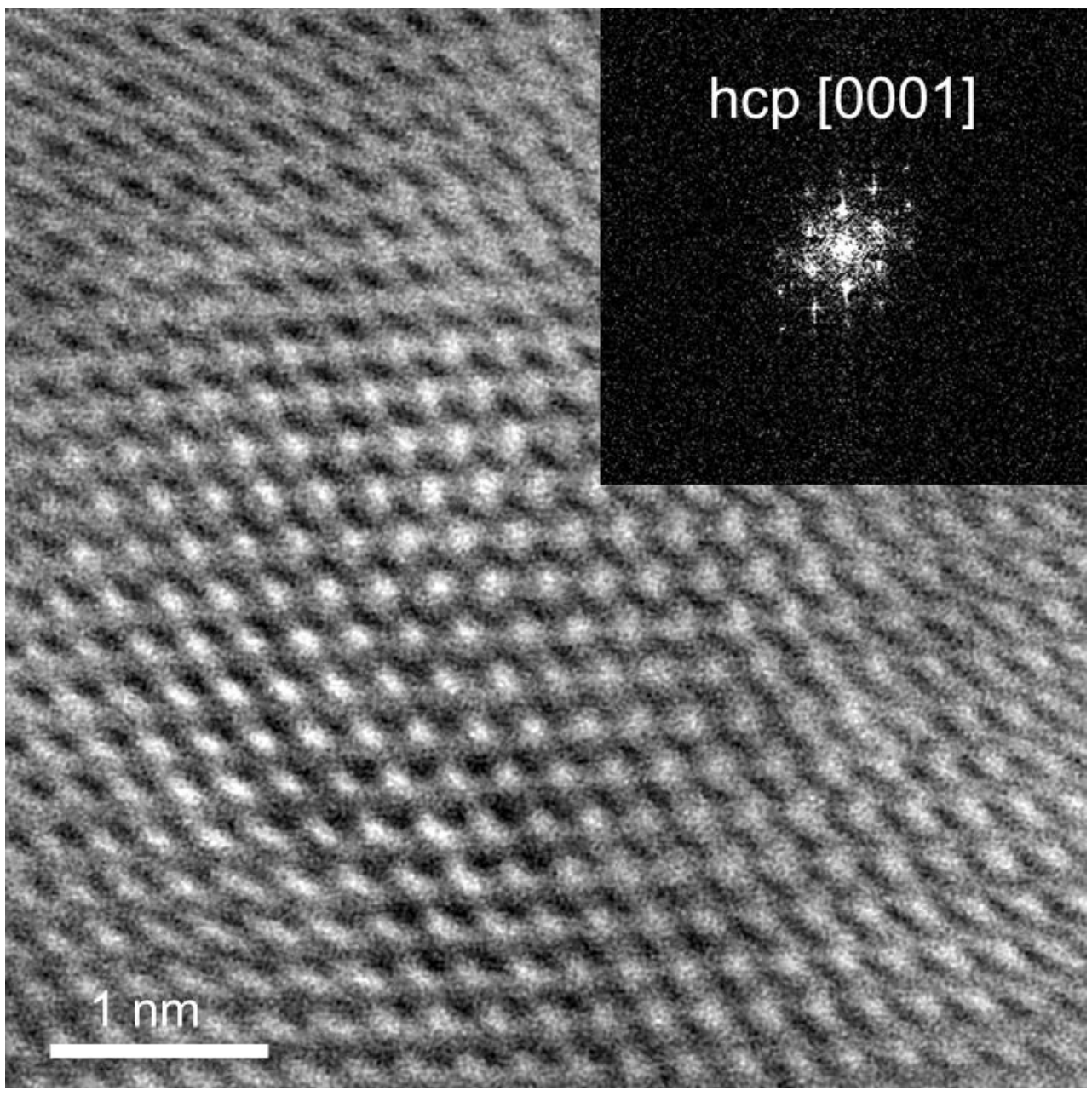

**Extended Data Fig. 2 | Atomic-resolution HAADF-STEM image and FFT pattern of the upper equiaxed grain in Fig. 4b.** This HAADF-STEM image was processed with the ABSF for the visibility of atomic columns.

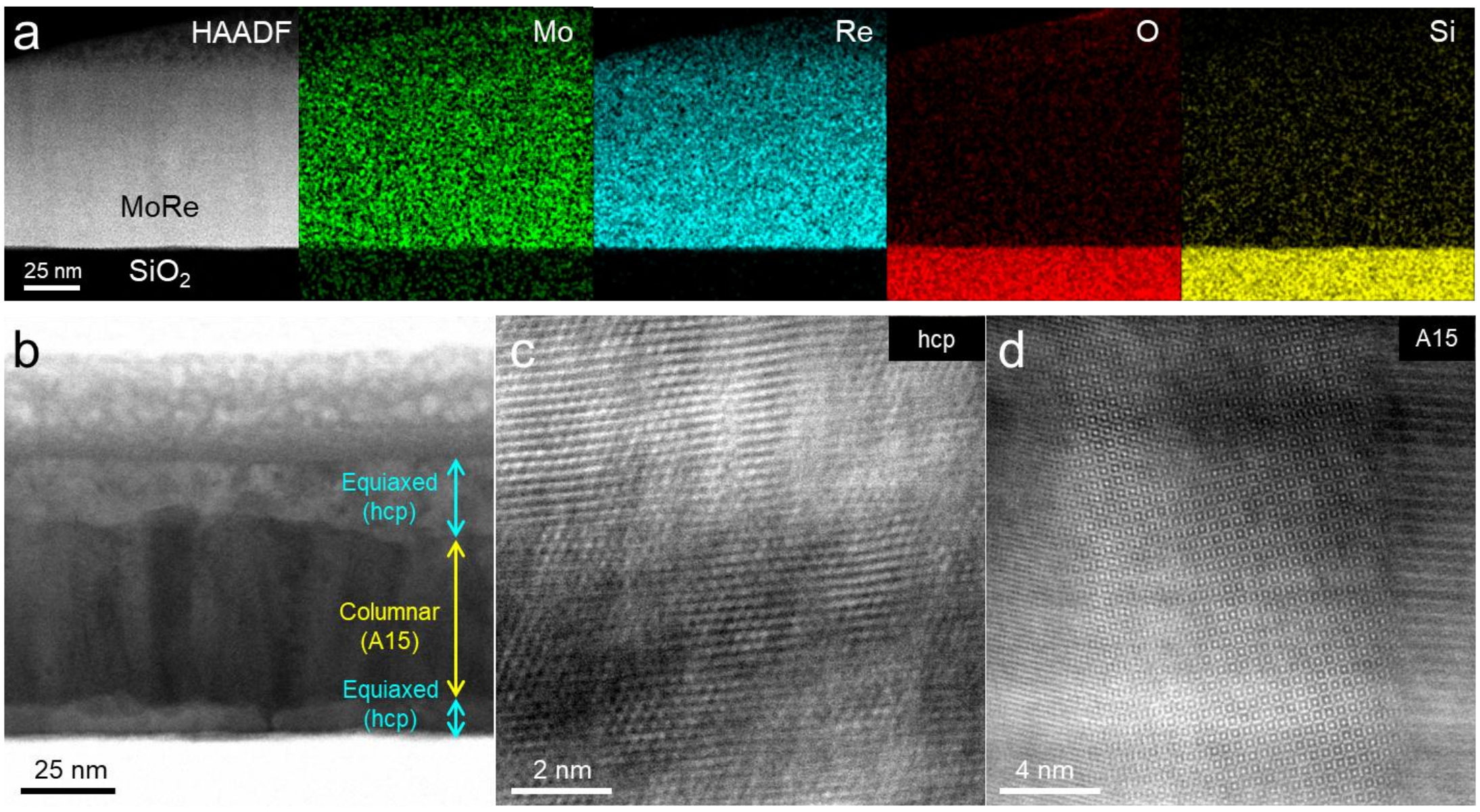

**Extended Data Fig. 3 | STEM observations of the MoRe annealed at 500°C. a** Cross-sectional HAADF-STEM image of the MoRe on $SiO_2$ annealed at 500°C and EDS mapping images for Mo, Re, O, and Si. **b** ABF image of the MoRe annealed at 500°C**. c,d** Atomic-resolution HAADF-STEM image of the hcp (c) and A15 (d) regions in **b**. The images of **c** and **d** were processed with the ABSF for the visibility of atomic columns.

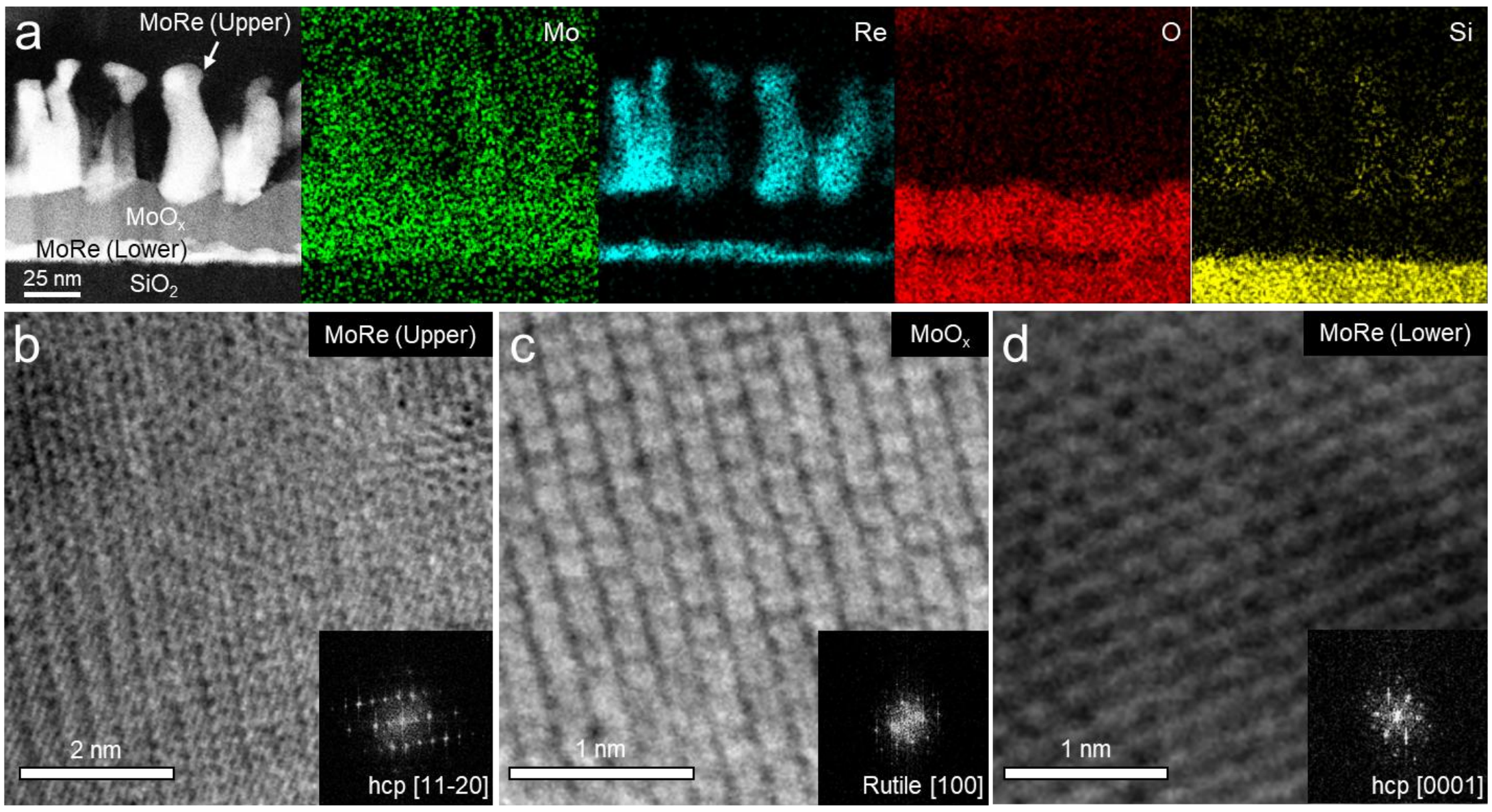

**Extended Data Fig. 4 | STEM observations of the MoRe annealed at 900°C. a** Cross-sectional HAADF-STEM image of the MoRe on $SiO_2$ annealed at 900°C and EDS mapping images for Mo, Re, O, and Si. **b** ABF image of the MoRe annealed at 900°C**. b–d** Atomic-resolution HAADF-STEM images of the (b) MoRe (Upper), (c) $MoO_x$ and (d) MoRe (Lower) regions in **a**. Insets in **b–d** show the corresponding FFT patterns.

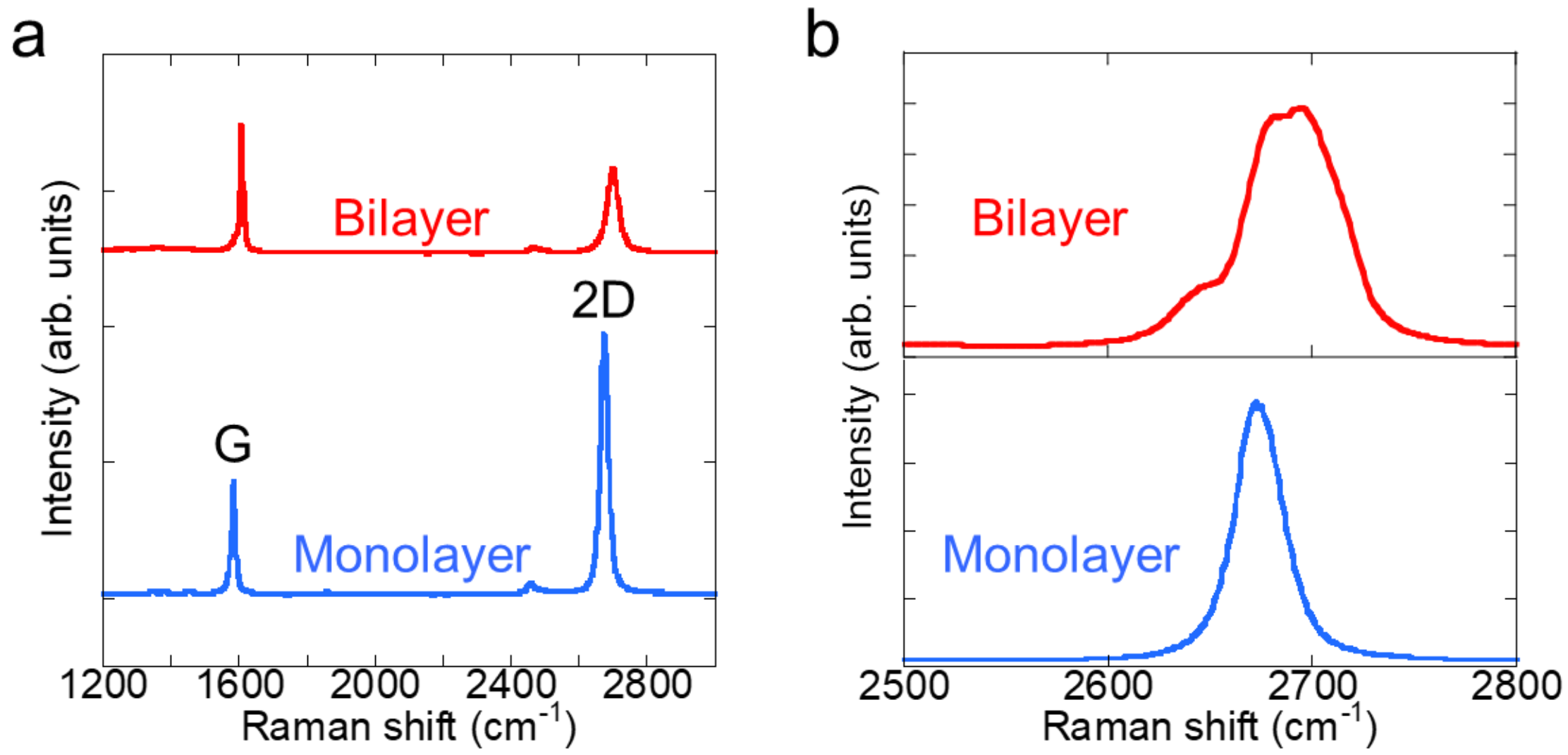

**Extended Data Fig. 5 | Raman characterization of monolayer and bilayer graphene. a** Raman spectra of monolayer and bilayer graphene. **b** Enlarged Raman spectra of the 2D peaks of monolayer and bilayer graphene.